\documentclass[conference]{IEEEtran}

\usepackage[cmex10]{amsmath}

\usepackage{array}

\usepackage[tight,footnotesize]{subfigure}

\usepackage{times,amsmath,epsfig}
\usepackage[T1]{fontenc}
\usepackage{graphicx} 
\usepackage{subfigure}
\usepackage{float}
\usepackage{boxedminipage}
\usepackage{enumerate}
\usepackage{color}
\usepackage{cite}
\usepackage{caption}

\newcommand{\ie}{\emph{i.e., }}
\newcommand{\eg}{\emph{e.g., }}

\begin{document}

\title{Identifying Randomly Activated Users via Sign-Compute-Resolve on Graphs}

\author{\IEEEauthorblockN{\v Cedomir Stefanovi\' c\IEEEauthorrefmark{1}, Dejan Vukobratovi\' c\IEEEauthorrefmark{2}, Jasper Goseling\IEEEauthorrefmark{3}, Petar Popovski\IEEEauthorrefmark{1}\\}
\IEEEauthorblockA{\IEEEauthorrefmark{1}Department of Electronic Systems, Aalborg University, Aalborg, Denmark, 
Email: \{cs,petarp\}@es.aau.dk \\
\IEEEauthorrefmark{2}Department of Power, Electronics and Communication Engineering, University of Novi Sad, Serbia, Email: dejanv@uns.ac.rs \\
\IEEEauthorrefmark{3}Stochastic Operations Research, University of Twente, The Netherlands, Email: j.goseling@utwente.nl
}
}

\maketitle

\begin{abstract}

In this paper we treat the problem of identification of a subset of active users in a set of a large number of potentially active users.
The users from the subset are activated randomly, such that the access point (AP) does not know the subset or its size a priori. 
The active users are contending to report their activity to the AP over a multiple access channel.
We devise a contention algorithm that assumes a combination of physical-layer network coding and $K$-out-of-$N$ signature coding, allowing for multiple detection of up to $K$ users at the access point.
In addition, we rely on the principles of coded slotted ALOHA (CSA) and use of successive interference cancellation to enable subsequent resolution of the collisions that originally featured more than $K$ users.
The objective is to identify the subset of active users such that the target performance, \eg probability of active user resolution and/or throughput is reached, which implies that the duration of the contention period is also not known a priori.
In contrast to standard CSA approaches, in the proposed algorithm each user, active or not, has a predefined schedule of slots in which it sends its signature. 
We analyze the performance of the proposed algorithm both in the asymptotic and non-asymptotic settings.
We also derive an estimator that, based on the observation of collision multiplicities, estimates how many users are active and thereby enables tuning of the length of the contention period.

\end{abstract}

\section{Introduction}

In wireless communications there are often scenarios in which a large number of terminals (\ie users) are associated with the same access point.
However, at a given time, the subset of users that is active and has data to send is random.
This requires a random access protocol for identification of the subset of active users.
Such protocol can be used, \eg to schedule the users after the identification procedure, such that each user can deliver its data to the access point.
Alternatively, if each user has a small amount of data to send, this can be done using the random access protocol. 

Slotted ALOHA (SA) \cite{R1975} is a widely used random access solution that offers simple implementation, but suffers from low throughput.
Namely, a typical SA premise is that collisions of two or more user packets in a slot are destructive, posing the key limitation on the performance.
However, recent advances on physical layer have given rise to motivation to relax or even discard this assumption.
Particularly, the recently proposed coded slotted ALOHA (CSA) \cite{PSLP2014} exploits successive interference cancellation to resolve collisions, providing for expected throughputs that asymptotically tend to 1 -- the upper limit on performance in absence of capture effect and/or multiple user detection.
Another recent advance is the use of physical-layer network coding (PLNC) for random access \cite{censor2012bounded,goseling14massap}, transforming the collisions into the (noiseless) sums of colliding packets.
In particular, if $K$-out-of-$N$ signature coding for multiple access\cite{MACbook} is combined with PLNC~\cite{goseling14massap,GSP2014,GSP2015}, sums of up to $K$ packets can be directly resolved.

In this paper we assume a scenario in which a subset of users is activated in a batch, such that neither the identities nor the number of active users in the subset are known.
We propose a random access algorithm that exploits combination of CSA and sign-compute-resolve frameworks \cite{GSP2014} in order to identify the active users.
The batch arrival is resolved within a contention period, consisting of equal-length slots, whose start and termination are signaled by the AP.
On slot level the contention is based on the principles of sign-compute-resolve, briefly introduced as follows.
The active users contend with packets that are PLNC-encoded representation of their signatures, where the signatures are unique representations of the user identities.
If there are up to $K$ user packets colliding in a slot, their PLNC-decoded sum is immediately resolvable into constituent signatures.
If there are more than $K$ packets colliding in a slot, the PLNC-decoded sum can not be resolved and is stored for later use.
Further, the active users transmit their PLNC encoded signatures in multiple slots, according to the pseudorandom transmissions schedules.
These schedules are predefined and there is one-to-one mapping from the signature of each user (active or not) to the specific transmission schedule. 
A resolution of a signature in any of the slots, enables its removal, \ie cancellation, from all other slots in which it appears and which contain more than $K$ colliding packets.
In this way, previously unresolved sums may become resolvable into constituent signatures, instigating new iterations of signature cancellations and resolutions; this is a founding feature of CSA methods.
The above described random access scheme is executed until the AP decides that the target performance is reached, which may be when the predefined fraction of users has been resolved and/or throughput is maximized.
In this respect, as the number of active users is unknown, the AP has to obtain its estimate along with the resolution of their signatures in order to decide whether the target performance has been reached. 

The pivotal goal of the paper is to analyze the proposed random access method and assess its potential.
We also evaluate the proposed method both in the asymptotic and non-asymptotic settings, for a simple instance of the scheme in which the transmission schedules of the users are defined such that each slot has the same number of potentially active users.
We also develop an estimator that fits naturally in the proposed framework, allowing for the termination of the scheme that adapts to the actual number of active users. 

The rest of the text is organized as follows.
In Section~\ref{sec:background} we provide a brief account of the related work.
System model is presented in Section~\ref{sec:model}.
The analysis of the asymptotic performance and the derivation of the proposed estimator is provided in Section~\ref{sec:analysis}.
The evaluation of the proposed scheme is performed in Section~\ref{sec:evaluation}.
Section~\ref{sec:conclusion} concludes the paper.

\section{Related Work}
\label{sec:background}

Coded slotted ALOHA, \ie codes-on-graphs based design and optimization of slotted ALOHA with successive interference cancellation (SIC) was established in \cite{L2011}.
This initial work was followed by application of different coding-for-erasures methods in CSA framework \cite{SP2013,Liva2015:CodedAloha}.
A CSA variant that exploits multiple user detection in a setting similar to ours was addressed in \cite{GS2013}.
The typical assumptions in these papers, as well as in CSA framework in general \cite{PSLP2014}, is that the number of the active users is a priori known, such that the parameters of the scheme, like frame lengths and probability distributions that control transmission schedules of the users, can be optimized with respect to it.
In contrast, the random access strategy proposed in this paper assumes that only the probability distribution governing the number of active users is known.
Another important difference to standard CSA is that the active users in standard CSA independently and uncoordinatedly establish their transmission schedules, whereas in this paper the transmission schedules of all users are determined by the AP, but actually followed only by the active users that are a priori unknown.
In terms of codes-on-graphs, we consider a design of right-node degree-distribution in rateless coding framework \cite{L2002} for up to $K$-out-of-$N$ decoding scenario, where only an a priori unknown fraction of the input symbols (\ie users) and the corresponding edges in the graph actually exist (\ie are active).
On the other hand, the standard CSA considers a decentralized design of left-node degree-distribution when the number of active users is a priori known.

Combination of PLNC and signature coding in SA framework was considered in \cite{goseling14massap,GSP2015} and in tree-splitting framework in \cite{GSP2014}.
Unlike in \cite{goseling14massap,GSP2015}, we consider a SIC-enabled scenario, such that the random access strategy is optimized on the basis of the contention period.

\section{System model}
\label{sec:model}

We consider a scenario where there are $N$ users and $N$ is known by the AP.
The user population is homogenous, \ie all users have the same transmission capabilities and are considered equally important.
Time is slotted, the slots are grouped in contention periods, and the users are synchronized on the slot and on the contention period basis.
We consider a single contention period.
The length of the contention period $M$ is not specified a priori, but determined on the fly by the AP, which also signals its termination.

We consider a batch arrival resolution, where the probability of a user arriving, \ie being active at the start of the contention period, denoted by $p_A$, is a priori known.
We assume that
\begin{align}
\label{eq:p_A}
 p_A = \frac{\alpha}{N} \ll 1.
 \end{align}
The number of active users $N_A$ is a binomially distributed random variable with mean $E [ N_A ] = p_A N = \alpha$ and probability mass function (pmf)
\begin{align}
\label{eq:n_prior}
p ( n ) = \text{P} [ N_A = n ]  = { N \choose n } p_A^n ( 1 - p_A )^{N - n } \approx \frac{\alpha^n}{n!}e^{-\alpha},
\end{align}
for $0 \leq n \leq N$.
The rest of the $N - N_A$ users are inactive. 

\begin{figure}[t]
\begin{centering}
\includegraphics[width=0.7\columnwidth]{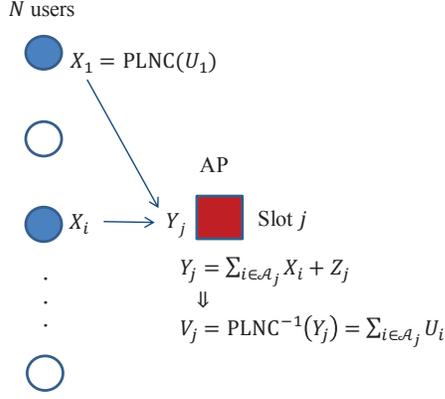}
\caption{Contention in slot $s_j$. Users active in $s_j$ are marked in blue. The AP a priori knows only the set of potentially active users in $s_j$, but not which of them are actually active.}
	\label{fig:contention}
\end{centering}
\vspace{-4mm}
\end{figure}

Each user has a unique signature $U_i$, which is a sequence of integers taking values in $\{0,\dots,q\}$. The length of these signatures is specified below.
The goal of the AP is to learn the set of active users through their signatures.
The users apply power control when transmitting, such that the multiple access channel (MAC) they experience can be modeled as a Gaussian MAC with unit channel gains.
Further, the active users also apply PLNC, \ie user $i$ transmits $X_i = \text{PLNC} ( U_i )$.
The AP observes in slot $j$ (henceforth denoted as $s_j$)
\begin{align}
Y_j = \sum_{i \in \mathcal{A}_j} X_i + Z_j, \; j = 1 \dots M,
\end{align}
where $\mathcal{A}_j$ is the set of active users that transmitted in slot $j$ and $Z_j$ is additive white Gaussian noise with unit variance.
We assume the use of reliable PLNC \cite{nazer2011procieee}, where the rates are selected in a way that the sum of the interfering signals i decoded reliably and the noise is effectively eliminated, such that the AP obtains
\begin{align}
\label{eq:signature_sum}
V_j = \sum_{i \in \mathcal{A}_j} U_i,
\end{align}
and the MAC channel effectively becomes an integer adder channel.\footnote{Note that PLNC is commonly specified in terms of operations in the finite field $\mathcal{F}_q$.  We adopt the framework of~\cite{GSP2015} in which the connection to integer operations is made precise and refer the reader to~\cite{GSP2015} for further details.}
This is depicted in Fig.~\ref{fig:contention}.

The user signatures are codewords of a $K$-out-of-$N$ signature code for the integer adder channel \cite{MACbook,GSP2014}, such that the AP is able to unambiguously, \ie uniquely resolve $V_j$ into constituent signatures if $| \mathcal{A}_j | \leq K $.
In \cite{MACbook,GSP2014} it was shown that the signatures can be constructed such that their length is $ L  \approx K \log N$ bits.
The use of signature coding allows for the generalization of the concept of the collision, as now the AP is able to directly exploit the slots in which there are $K$ or less active users.
In essence, joint application of PLNC and signature coding effectively provides an error-free multiple user detection of up to $K$ users at the AP.

Finally, we assume that the AP is capable of storing unresolved sums $V_j$ and performing SIC, through which the sums that contain more than $K$ signatures of the active users may become useful in later stages. 

\subsection{Access Strategy}

Our interest is to design an access strategy that, given the uncertainty in the set of the active users and knowing only $p_A$ and $N$, achieves favorable performance in the terms of the probability of identifying active/inactive users and/or throughput (both parameters are formally defined later).
We adopt an approach in which the transmission schedule of each user is fixed at the beginning of the contention period by the AP and signaled to the users.
This could be achieved, \eg via beacon by which the AP signals the start of the contention period and informs the users about their transmission schedules.\footnote{This could be effectively done by broadcasting a common seed, which in combination with the user signatures, determines the transmission schedules.}
The transmission schedule effectively applies to active users only and defines in which slots of the contention period the users should transmit their signatures.
The design of the schedules is crucial for the achievement of an effective resolution strategy, when both the set and the size of the set of active users is a priori not known.

For the design of the schedules, we adopt an approach inspired by the rateless coding paradigm \cite{L2002,SPV2012}.
Specifically, the pivotal design parameter of the considered approach is the specification of the number of potentially active users in all slots $s_j$, denoted as slot degrees $|s_j|$; this is done according to a predefined slot degree distribution
\begin{align}
\Omega_{d_S} & = \text{P} [ |s_j| = d_S ], \; 1 \leq d_S \leq N, \text{ and }  \sum_{d_S = 1}^{N} \Omega_{d_S} = 1.
\end{align}
Observe that slot degrees are independent identically distributed (i.i.d) random variables, with the expected value $\beta = \text{E} [ | s | ]$.
Further, after specifying $| s_j | $, the AP schedules $|s_j|$ users in slot $s_j$ by choosing them uniformly randomly from the set of all users (comprising both active and inactive users).
Therefore, only a random subset of users $\mathcal{A}_j$  from $|s_j|$ scheduled ones is actually present in the slot, corresponding to the active users.
Assume that $|s_j| = d_S$ and $N_A = n $.
The number of active users $| \mathcal{A}_j |$ is a random variable in the range $[0, d_S ]$, whose conditional pmf is
\begin{align}
\label{eq:d_A}
 p  ( d_A | d_S, n ) & = \text{P} [ |\mathcal{A}_j| = d_A | d_S , n ] \nonumber \\
			  & = \begin{cases} 
			   { n \choose  d_A } { N - n \choose { d_S - d_A } } / { N \choose d_S}, & n \geq d_A \\ 
			   0, &  n < d_A
			   \end{cases},
\end{align}
which, for fixed $d_S$ and when $n,N \rightarrow \infty$, can be approximated  as
\begin{align}
p  ( d_A | d_S, n ) \approx { d_S \choose d_A } \left(  \frac{n}{N} \right)^{d_A} \left(1 - \frac{n}{N} \right)^{d_S - d_A},
\end{align}
The active-degree distribution of a slot is
\begin{align}
\Psi_{d_A} & = \text{ P} [ |A_j| = d_A ] = \sum_{n=1}^{N} \sum_{d_S = 1}^{N} p ( d_A | d_S, n ) \Omega_{d_S} p ( n ), 
\end{align}
for $0 \leq d_A \leq N$.

In order to avoid non-unique resolvability of the signature sums when $|\mathcal{A}_j| > K$, one could set $\max{|s_j|} = K$, $ \forall j$.
However, such an approach is too restrictive and it is beneficial to allow $\max(|s_j|) > K $, as shown in later sections.
In order to cope with the case when it happens that $|\mathcal{A}_j| > K$, we assume that the AP is able to infer how many signatures are actually present in the sum $V_j$.
This could be done, \eg using an indicator symbol that is set to 1 and located at a fixed and the same position in every signature, such that the sum of these symbols indicates the value of $|A_j|$ \cite{MACbook, GSP2014}.
We further comment on this issue in Section~\ref{sec:estimation}. 

From the user perspective, the assigned schedule defines:
\begin{enumerate}[(i)]
\item how many times a user sends a replica of its signature, which is denoted as the user degree $|u_i |$, $ 1 \leq i \leq N $,
\item slots in which these replicas occur.
\end{enumerate}
It can be shown that the user degrees are characterized by the user degree distribution
\begin{align}
\text{P} [ | u_i | = d_U ] & = \Lambda_{d_U} \approx { M \beta \choose d_U } \left( \frac{1}{N} \right)^{d_U} \left( 1 - \frac{1}{N} \right)^{M \beta - d_U} \\
& = \frac{\left( \frac{M }{N} \beta \right)^{d_U}}{d_U !} e^{-\frac{M}{N} \beta}, \; 1 \leq d_U \leq M,
\end{align}
where $M$ is the length of the contention period that is adaptively determined, $ M \beta $ is the total number of signature transmissions taking place in the contention period, and the expected user degree is $\text{ E } [ | u | ]= \frac{M}{N} \beta$.


\subsection{Resolution of the Active Users Status}

The algorithm for the resolution of the user-activity status resembles the iterative belief-propagation erasure decoding of rateless codes \cite{L2002}, with a difference that up to $K$ simultaneous transmissions can be resolved.
Specifically, after each slot, the AP performs PLNC decoding, stores the received signature sum \eqref{eq:signature_sum} and attempts resolution of the involved signatures.
If the AP succeeds (\ie there were up to $K$ signatures in the sum), then it also removes the resolved signatures from all previously stored and unresolved signature sums, in which these signatures may be involved, \ie the AP performs successive interference cancellation.\footnote{Recall that the AP defines the schedules and thereby knows in which slots the signature replicas occur.}
In the next step, the AP attempts to resolve the signature sums that were affected by SIC in the previous step.
Note that, after resolution of a signature sum, the AP also implicitly learns which of the potentially involved signatures (\ie users) are inactive; however, this information does not help resolution of the active signatures in the proposed framework.

\subsection{Performance Parameters}
\label{sec:perf_par}

The basic performance parameter of interest is the conditional probability of resolving an active user, given that there are $N_A = n$ users active, 
\begin{align}
\label{eq:p_AR_Cond}
p_{R } (M  |  n ) = \frac{N_{R } (M | n )}{n}, 
\end{align}
where $N_{R} (M, n )$ 
is the number of the resolved active users during $M$ slots of the contention period, given that $N_A = n$.
Averaging over $N_A$ produces
\begin{align}
p_{R } ( M ) & = \sum_{n=1}^{N}  p_{R } (M  | n ) \, p ( n ) \approx \sum_{n=1}^{N}  \frac{N_{R} (M | n )}{n} \, \frac{\alpha^n}{n!}e^{-\alpha}, 
\end{align}

We are also interested in throughput, \ie the rate by which the AP resolves user activity status over slots.
The throughput conditioned on $N_A = n$ is 
\begin{align}
\label{eq:throughput_cond}
T( M | n )  & = \frac{N_{R} (M | n ) } { M K }= \frac{p_{R} (M  | n ) \, n } { M K },
\end{align}
and the average throughput is
\begin{align}
\label{eq:throughput}
& T ( M  ) = \sum_{n=1}^{N}  T (M  | n ) \, p ( n )  \approx \frac{1}{MK} \sum_{n=1}^{N}  p_{R} (M  | n  )  \frac{\alpha^n e^{-\alpha}}{( n - 1 )!}.
\end{align}
where the normalization through factor $K$ takes into account the overhead required by the signature coding.

\subsection{Graph Representation}

\begin{figure}[t]
\includegraphics[width=0.7\columnwidth]{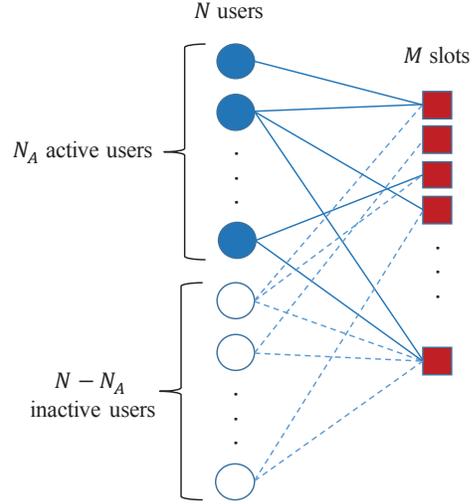}
\caption{Graph representation of the contention period.}
	\label{fig:graph}
\end{figure}

We represent the contention process through a graph, see Fig.~\ref{fig:graph}.
The nodes on the left represent users and the nodes on the right represent slots of the contention period.
There are two types of users, active - marked in blue, and inactive - marked in white.
In line with the description of the access strategy, all the edges of the graph are pre-determined before the contention period starts, corresponding to the individual schedules.
Among them, only the edges incident to the active users are actually realized and correspond to the signatures' transmissions; these edges are represented by solid lines.
The edges corresponding to the inactive users are represented by dashed lines.
Before the resolution starts, the AP does not know which edges in the graph are actually active/inactive.

\begin{figure*}[t]
\begin{centering}
\includegraphics[width=1.75\columnwidth]{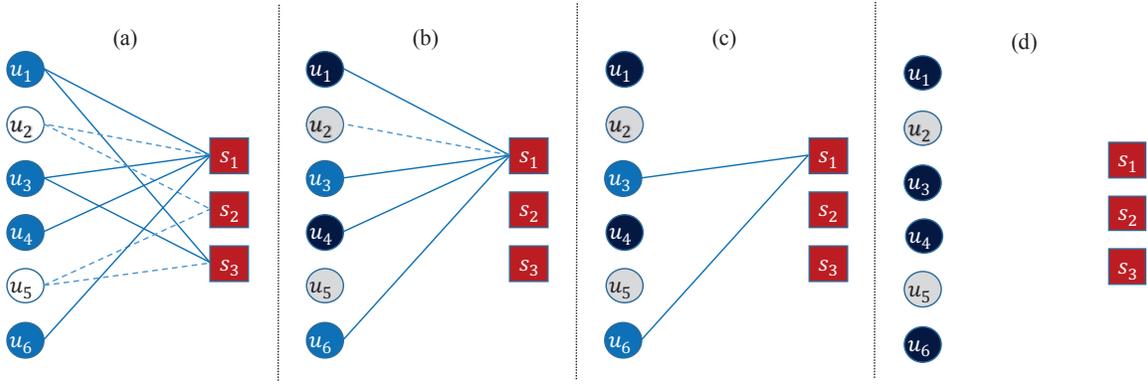}
\vspace{-2mm}
\caption{An example graph of user resolution for $2$-out-of-$N$ signature coding. (a) Contention outcome. (b) The signature sums in $s_2$ and $s_3$ are resolved, where $s_2$ contains no active signatures and $s_3$ contains sum of signatures $U_1$ and $U_4$. (c) Replicas of $U_1$ and $U_4$ are cancelled from $s_1$, leaving two active signatures in the corresponding sum. (c)  $U_3$ and $U_6$ are resolved in $s_1$.}
	\label{fig:decoding}
\end{centering}
\end{figure*}

The resolution algorithm, in terms of graph representation, can be seen as removal of the involved edges, see Fig.~\ref{fig:decoding}.
Specifically, when the signature sum is resolved, all edges incident to a slot are removed; this holds both for the active and inactive edges, as the latter do not contribute to the sum.
Finally, when an edge incident to a user node is removed, then all the other edges incident to the same user node are also removed; this corresponds to the SIC step of the algorithm.

\section{Analysis}
\label{sec:analysis}

\subsection{Asymptotic Probability of User Activity Status Resolution}

We derive the probability of user activity status resolution in asymptotic setting when $N \rightarrow \infty$, using the approach that is based on the and-or tree evaluation \cite{LMS1998}.
In order to foster a tractable analysis, we use the following approximation to \eqref{eq:throughput}
\begin{align}
T ( M ) \approx T ( M | \, \text{E} [ N_A ] ) = T ( M | \alpha),
\end{align}
where $\text{E} [ N_A ] = \alpha $, see \eqref{eq:p_A}.
In other words, we develop the analysis for the expected number of active users.
In further text, we drop the subscript referring to user/slot, when there is no ambiguity.

\begin{figure}[t]
\centering
\includegraphics[width=0.9\columnwidth]{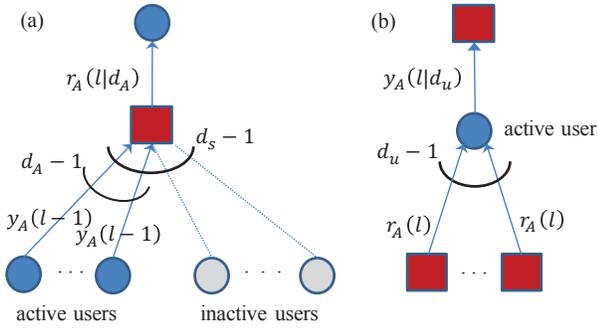}
\caption{(a) Slot and (b) user nodes in and-or tree evaluation.}
	\label{fig:and-or}
\end{figure}

We assume that the decoding algorithm is in its $l-$th iteration.
We start by considering a slot node, as depicted in Fig.~\ref{fig:and-or}(a), and assume that there are $d_A \geq 1 $ active edges. 
We consider one of the randomly chosen active edges that has not been removed in previous iterations and by $r_A ( l | d_A )$ denote the probability that the edge is \emph{not} removed in the current iteration. 
Finally, by $y_A ( l - 1 )$ we denote the probability that an active edge incident to the slot was \emph{not} removed in the previous iteration.\footnote{For a more detailed introduction to and-or tree evaluation, we refer the interested reader to \cite{LMS1998}.}

The probability that an active edge incident to a slot is not removed in the current iteration is:
\begin{align}
\label{eq:conditional_r_A}
& r_A  ( l | d_A ) =  1 - \nonumber \\ 
& - \sum_{h=1}^{\min ( d_A,K ) } { d_A - 1 \choose h - 1} y_A( l - 1 )^{h-1} ( 1 - y_A ( l - 1 ) )^{d_A - h}, 
\end{align}
for $d_A \geq 1 $, which is due to the properties of the $K$-out-of-$N$ signature coding.
On the other hand, if any of the edges incident to a user node with degree $d_U$ becomes removed, Fig.~\ref{fig:and-or}(b), all other edges are also removed in the SIC step.
Using analogous notation, this is expressed as
\begin{align}
\label{eq:conditional_y}
y_A ( l | d_U ) = r_A ( l  )^{d_U - 1}.
\end{align}

In order to derive unconditional probabilities of not removing an edge incident to a slot/user node, we need \emph{edge-oriented} degree distribution \cite{LMS1998}, \ie the probabilities that an edge is incident to a slot/user node of a certain degree; these are
\begin{align}
\label{eq:gamma}
\psi_{d_A} = \frac{d_A  \, \Psi_{d_A}}{ \sum_j \Psi_j } , d_A \geq 1, \text{ and } \lambda_{d_U} = \frac{ d_U  \, \Lambda_{d_U} }{ \sum_i \Lambda_i  }, d_U \geq 1.
\end{align}
The probability that an active edge incident to a slot node is not removed in iteration $l$ is
\begin{align}
\label{eq:r_A_l}
r_A ( l ) & = \sum_{d_A=1}^{\infty} \psi_{d_A}  r_A ( l | d_A ), 
\end{align}
The probability that an edge incident to a user node is not removed in iteration $l$ is
\begin{align}
y_A ( l  ) & = \sum_{d_u} \lambda_{d_U} \, y_A ( l | d_U ) = \sum_{d_U} \lambda_{d_U} \, r_A ( l )^{d_U - 1} \\
& \approx e^{- ( 1 + \epsilon ) \beta ( 1 - r_A ( l  ))}, \label{eq:y_A_l}
\end{align}
where $(1 + \epsilon) = \frac{M}{N}$.

The and-or tree evaluation starts with the initial value $y_A ( 0  ) = 1$, and outputs the probability of user resolution:
\begin{align}
p_{R} ( M | \alpha ) = 1 - \lim_{l \rightarrow \infty} y_{A} ( l ), 
\end{align}
from which the asymptotic throughput can be derived as
\begin{align}
T ( M | \alpha) = \frac{p_{R} ( M | \alpha ) \,  p_A}{ ( 1 + \epsilon )K } .
\end{align}

\subsection{Estimating Number of Active Users}
\label{sec:estimation}

In order to keep track of the probability of the user resolution, the AP has also to learn the value of $N_A$, \ie how many users are active.
A way to estimate $N_A$ is to use the information on how many active users $|A_j|$ are involved in the sum $V_j$.
As elaborated in Section~\ref{sec:model}, we assume that every signature contains an indicator symbol, the sum of which informs the AP of the number of the active signatures $|\mathcal{A}_j|$ involved in $V_j$, $j = 1, \dots, M$, see \eqref{eq:signature_sum}.
Nevertheless, due to the operating range of the AP receiver, it is sensible to assume that there is an upper limit on $|\mathcal{A}_j|$, denoted by $K_{\max}$, after which the AP only knows that the number of active signatures is equal to or larger than $K_{\max}$.
We consider only the case when $K_{\max} \geq K$, for obvious reasons.
In further text, we propose a simple maximum a posteriori (MAP) estimator that makes use of the observations of $|\mathcal{A}_j|$, $ 1 \leq j $.

The a priori probability that the number of active users $N_A$ is equal to $n$ is given by \eqref{eq:n_prior}.
Assume for the time being that the exact value of $|\mathcal{A}_j|$ can be detected.
With a slight abuse of notation, where we denote by $|s_j|$ and $|\mathcal{A}_j|$ the realizations of the slot degree and the number of the active users in slot $s_j$, the a posteriori probability that $N_A = n$ is
\begin{align}
\label{eq:a-posteriori}
& p ( n \, \boldsymbol{|} \,  | \mathcal{A}_1 |, \dots , |\mathcal{A}_m | ;  |s_1|, \dots, |s_m| )  = \nonumber \\
 & =  \frac{ p ( | \mathcal{A}_1 |, \dots , |\mathcal{A}_m | \, \boldsymbol{|} \, | s_1 |,  \dots, |s_m| ;  n ) \, p ( n ) }{ p ( | \mathcal{A}_1 |, \dots , | \mathcal{A}_m | \, \boldsymbol{|} \, | s_1 |,  \dots, |s_m| ) } \\
 & =  \frac{ \prod_{j=1}^{m} p ( | \mathcal{A}_j | \, \boldsymbol{|} \, | s_j | , n ) \, p ( n ) }{ \prod_{j=1}^{m} p ( | \mathcal{A}_j | \, \boldsymbol{|} \, | s_j | ) },
\end{align}
where $m$ denotes the current number of the observed slots and where we used the fact that the selection of the slot degrees and of the users is independent over slots.
The MAP estimator provides $n$ that maximizes \eqref{eq:a-posteriori}, \ie 
\begin{align}
\arg \max_n \{ p ( n  \, \boldsymbol{|} \,  | \mathcal{A}_1 |, \dots , | \mathcal{A}_m | ; |s_1|, \dots, & |s_m| ) \},
\end{align}
or, equivalently,
\begin{align}
\arg \max_n \{ \log ( p  ( n \, \boldsymbol{|} \,  | \mathcal{A}_1 |, \dots , | \mathcal{A}_m | ; |s_1|, \dots, & |s_m| ) ) \}.
\end{align}
Further, we have
\begin{align}
& \log  ( p  (n \, \boldsymbol{|} \,  | A_1 |, \dots , | A_m| ; |s_1|, \dots, |s_m| ) ) =  \log (  p ( n ) ) + \nonumber \\
& + \sum_{j=1}^m \log (p ( | A_j | \, \boldsymbol{|} \, | s_j | , n )  )  - \sum_{j=1}^m \log ( p ( | A_j | \, \boldsymbol{|} \, | s_j | ) ) .
\label{eq:aux}
\end{align}
Instead of maximizing \eqref{eq:aux}, it is equivalent to maximize
\begin{align}
\label{eq:max}
F ( m, n ) = & \sum_{j=1}^m \log ( p ( | \mathcal{A}_j | \, \boldsymbol{|} \, | s_j | , n )  ) + \log ( p ( n ) ). 
\end{align} 
The maximum can be obtained by solving the equation
\begin{align}
\label{eq:maximization}
\frac{\partial F ( m, n ) }{\partial n} = & \sum_{j=1}^m \left\{ \frac{| \mathcal{A}_j |}{ n } -  \frac{ |s_j| - | \mathcal{A}_j | }{N - n } \right\}  + \nonumber \\
& + \log \alpha - \sum_{h=1}^n \frac{1}{h} + \gamma = 0,
\end{align}
where $\gamma$ is the Euler--Mascheroni constant, and where we substituted \eqref{eq:d_A} and \eqref{eq:n_prior} into \eqref{eq:max}.

The estimation is performed after each observed slot, upon execution of the signature resolution algorithm.
Specifically, after each observed slot $s_j$, the AP detects whether $| \mathcal{A}_j | \leq K $.
If yes, the AP resolves the sum $V_j$ and iterates SIC and signature resolution step, as described in Section~\ref{sec:model}.
After the signature resolution, the AP obtains up-to-date information on all $\mathcal{A}_j$, $1 \leq j \leq m$,  which may include learning the exact values of some $| \mathcal{A}_j |$ for which the AP previously only knew that $| \mathcal{A}_j | \geq K_{\max}$.
Using all $|A_j|$ that are exactly known, the AP solves \eqref{eq:maximization} and obtains the current estimate of $N_A$.
Obviously, as the slots of the contention period elapse, the estimate becomes more refined.
We note that disregarding the slots for which it remains only known that $|\mathcal{A} | \geq K_{\max}$ is clearly a suboptimal solution, but it allows for an analytically elegant estimator \eqref{eq:maximization}.

\section{Evaluation}
\label{sec:evaluation}

\begin{figure}[t]
        \centering
        \subfigure[Maximal probability of active user resolution $p_R^*$ and its upper bound $p_U$.]
        {
                \includegraphics[width=0.85\columnwidth]{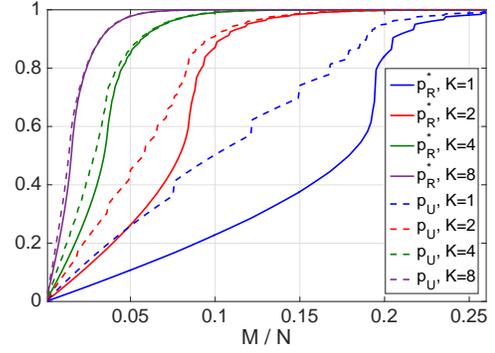}
                \label{fig:p_R}
     	}        
        \subfigure[Maximal asymptotic throughput $T^*$.]
        {
                \includegraphics[width=0.85\columnwidth]{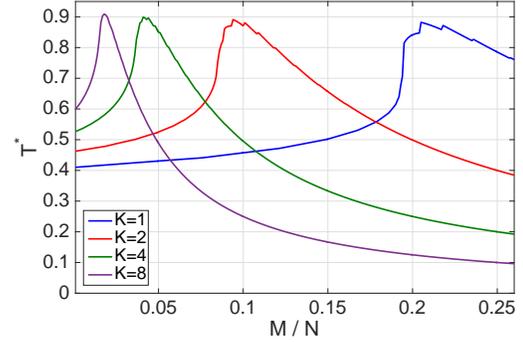}
                \label{fig:T}
        }  
        \subfigure[Optimal slot degree $\beta^*$.]
        {
                \includegraphics[width=0.85\columnwidth]{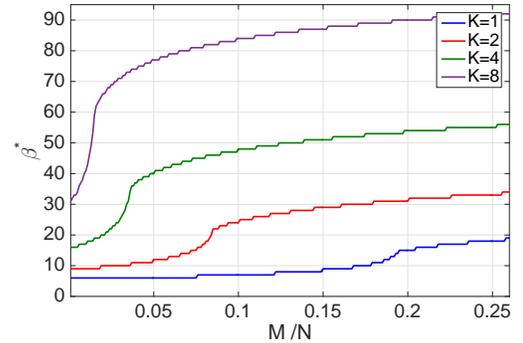}
                \label{fig:beta}
        }
        \vspace{-1mm}    
   \caption{Asymptotic performance, $p_A = 0.2$.}\label{fig:asymptotic}  
\end{figure} 

As an illustration, we evaluate the performance of the proposed scheme in a simplified instance in which all slot degrees $|s_j|$ are equal and set to $\beta$, \ie
\begin{align}
\Omega_{d_s} = \begin{cases}
1, & \, d_s = \beta, \\
0, & \, \text{otherwise}.
\end{cases}
\end{align}
We do not investigate the optimality of such an approach in the paper, and the design of more involved slot degree-distributions which may offer improved performance is part of our ongoing work.
In the rest of the text we assume that the probability of user activation is $p_A = 0.2$.

\subsection{Asymptotic Performance}

In the asymptotic case, when $N \rightarrow \infty$, it can be shown that \eqref{eq:gamma} becomes
\begin{align}
\psi_{d_A} = { \beta - 1 \choose d_A - 1 } p_A^{ d_A - 1 } ( 1 - p_A )^{\beta - d_A }, \, d_A \geq 1,
\end{align}
which, combined with \eqref{eq:r_A_l} and \eqref{eq:y_A_l}, enables the evalution.

Fig.~\ref{fig:asymptotic} shows the asymptotic performance for constant $s_j$.
In particular, the figure depicts (a) the maximal probability of resolution $p_R^*$, (b) the corresponding maximal expected throughput $T^*$, and (c) the corresponding optimal slot degree $\beta^*$ for which $p_R^*$ is achieved, as functions of the ratio $\frac{M}{N}$, \ie the number of slots in the contention period vs the total number of users.
Obviously, Fig.~\ref{fig:p_R} shows that increase in $K$ allows for faster increase in $p_R^*$.
Also, all the curves corresponding to $p_R^*$ show a threshold-like behavior characteristic for CSA~\cite{PSLP2014}, \ie there is a steep increase at a certain value of $M/N$ when the SIC gain emerges.
This steep increase in $p_R^*$ happens at $\frac{M}{N} \approx \frac{0.2}{K} = \frac{p_A}{K}$; note that $M = \frac{p_A N }{ K}$ is the minimum number of slots required to resolve $N_A= p_A N $ users in the proposed scenario.
Fig.~\ref{fig:p_R} also depicts the upper limit on the probability of the active user resolution $p_{U} = 1 - \text{P} [ | u_i | = 0 ]  \approx 1 -  e^{ - \frac{M \beta^* }{ N } }$,
\ie the upper bound assumes that all users that transmitted at least once become resolved.
Obviously, as $K$ grows, the gap between $p_R^*$ and $p_U$ narrows.
Fig.~\ref{fig:T} shows that $T^*$ is maximized for $M = \frac{p_A N }{ K}$, and that the maximum values of $T^*$ slowly increase with $K$.
This means that investing in signature coding, \ie increasing $K$, does not notably impact throughput in this simple scheme, but does benefit attaining the upper bound on the $p_R$ performance.
Finally, Fig.~\ref{fig:beta} shows that $\beta^*$ increases with $K$, which could be expected.

\begin{table}[t]
\begin{center}
\begin{tabular}{|c|| c| c| c| c| c| c |} \hline
\rule{0pt}{2.5ex} 
$K$ & $ \bar{f}_{R,E}$ & $ \bar{f}_{R,A}$ & $ \overline{T}^* $ & $ \overline{\Delta n_E} $ & $ \overline{|\Delta n_E|} $ & $\beta^*$
\\ \hline \hline
1 & 0.92  & 0.92 & 0.82 & 0.01  & 0.05 & 15 \\ \hline
2 & 0.82 &  0.82 & 0.83 & -0.01 & 0.05 & 21 \\ \hline
4 & 0.78 & 0.77 & 0.78 & -0.01 & 0.05 & 33 \\ \hline
8 & 0.73 &  0.73 &  0.78 & 0.01 & 0.05 & 54 \\ \hline
\end{tabular}
\caption{Performance for $p_A = 0.2$, $N=1000$, $H = 0.7$.}
\label{tab:results}
\vspace{-4mm}
\end{center}
\end{table}

\subsection{Non-asymptotic Performance}

Non-asymptotic performance is evaluated via simulations for $N=1000$, $K_{\max} = 10$ and varying $\beta$.
The aim is to maximize the expected throughput, \ie to find the optimal $\beta$, when the estimated fraction of resolved (active) users $f_{R,E}$ is above a predefined threshold $H$.\footnote{This is just one way to optimize the scheme; one could also consider a target estimation accuracy, or maximize expected throughput and disregard $H$, etc.
Such exploration of the design space exceeds the scope of this work.}
In each simulation run, the slots are added and user resolution and estimation executed until $f_{R,E}=\frac{N_R}{N_E} \geq H$, where $N_R$ and $N_E$ are the current number of resolved active users and the current estimate of the total number of active users, respectively.
When $f_{R,E}\geq H$, then $N_R$, $N_E$, the actual number of active users $N_A$ and the length of the contention period $M$ are recorded.

Table~\ref{tab:results} presents performance parameters for $H = 0.7$: (i) $f_{R,E}$, (ii) the actual fraction of resolved users $f_{R,A} = \frac{N_R}{N_A}$, (iii) the maximal throughput $T^* = \frac{N_R}{M}$, (iv) the relative estimation error $ \Delta n_E = \frac{ N_E - N_A }{ N_A}$, and (v) the relative absolute error $ | \Delta n_E | = \frac{|N_E - N_A |}{ N_A} $, averaged over simulation runs for (vi) $\beta^*$ that maximizes the average throughput.
Obviously,  $\bar{f}_{R,E}$ is rather close to $\bar{f}_{R,A}$ and the estimation errors are small.
Further, $\bar{f}_{R,E}$ is actually considerably larger than $H$.
In particular, choice of $H=0.7$ corresponds to the value of $p_R^*$  in Fig.~\ref{fig:p_R} for which there is a sharp increase in $p_R^*$ as $\frac{M}{N}$ grows, which is due to SIC; this corresponds to a sudden jump in $f_{R,E} $ well above $H$ in a course of a single slot in the non-asymptotic scenario.\footnote{The same phenomenon was observed in \cite{SP2013}.}
The values of $\beta^*$ are quite close to the ones presented in Fig.~\ref{fig:beta}, for which the corresponding $p_R^*$ is equal to $\bar{f}_{R,A}$.
Finally, $\overline{T}^*$ is lower than in the asymptotic case, which can be expected in CSA-based schemes \cite{L2011,SP2013}.

\section{Concluding Remarks}
\label{sec:conclusion}

We presented a scheme for the resolution of batch arrival of the unknown subset of users, leveraging on sign-compute-resolve and CSA frameworks.
In contrast to standard CSA, we considered design of the slot-degree distributions when the number active users is a priori not known.
We developed analytical description of the asymptotic performance and the estimator of the number of active users, which is an integral part of the proposed scheme. 
We evaluated the scheme in a simple instance in which all slot degrees are the same.
Optimization of the slot-degree distribution and the criteria for the contention termination are part of our ongoing work.

\section*{Acknowledgment}

The research presented in this paper was supported in part by the Danish Council for Independent Research, Grant No. DFF-4005-00281, and in part by COST Action IC1104.

The authors would like to thank Marko Angjelichinoski for the software implementation of the estimator.

\end{document}